\documentstyle[12pt]{article}
\begin{document}
\begin{center}
{\large \bf 
Herd Behaviors in Financial Markets \\

\vspace*{.5in}

\normalsize 
Kyungsik Kim$^{a,*}$, Seong-Min Yoon$^{b}$, J. S. Choi$^{a}$
, Hideki Takayasu$^{c}$  

\vspace*{.1in}

{\em 
$^{a}$Department of Physics, Pukyong National University,\\
Pusan 608-737, Korea\\
$^{b}$Division of Economics, Pukyong National University,\\
 Pusan 608-737, Korea\\ 
$^{c}$Sony Computer Science Laboratories, 3-14-13 Higashi-gotanda, \\
Shinagawa-Ku 141-0022 Tokyo, Japan} \\
}
\end{center}

\hfill\\
%
%
%
\baselineskip 24pt

We investigate the herd behavior of returns for the yen-dollar exchange rate
in the Japanese financial market.
It is obtained that the probability distribution $P(R)$ of returns $R$ 
satisfies the power-law behavior
$P(R) \simeq R^{-\beta}$ with the exponents $ \beta=3.11$(the time interval $\tau=$ one minute) and $3.36$($\tau=$ one day).
The informational cascade regime appears in the herding parameter
$H\ge 2.33$ at $\tau=$ one minute,
%
while it occurs no herding at $\tau=$ one day.
Especially, we find that the distribution of normalized returns shows
a crossover to a Gaussian distribution at one time step
$\Delta t=1$ day.

%
\vskip 10mm 
\hfill
\vskip 1mm 
\hfill\\
$^{*}$Email: kskim@pknu.ac.kr; Tel.: +82-51-620-6354; Fax: +82-51-611-6357  \\

\newpage


Recently, the microscopic models in financial markets $[1-3]$  have received considerable attention.
Of many crucial models, the main concentration  is on 
the herding multiagent model $[4,5]$ and the related percolation models$[6,7]$, 
the democracy and dictatorship model $[8]$, the crowd- anticrowd theory,
the self-organized dynamical model $[9]$, the cut and paste model, and the fragmentation and
coagulation model $[10]$.
One of microscopic models in the self-organized phenomena
is the herding model $[11,12]$, in which some degrees of coordination
among a group of agents share the same information or the same rumor and
make a common decision in order to create and produce the returns.
Directly, we can discuss the dynamical herd bechavior via analyzing and simulating the real tick data. 
There are three important reasons to be influenced into the herd behavior $[13]$:
First, it exists the crash model that the herds may be occured by the biased information between investors.
Second, the return structure of fund managers may be sensitive to the herd behavior, since bank and stock company 
influence powerfully to investors. Lastly, fund manager and market analysist may play a crucial role
to essentially determine the investment, in order to maintain their reputation and credit. 
Particularly, 
it is of interest for the herd model to search for the bubbles and crashes
in econophysical system. 
The probability distribution of returns
shows a power-law behavior for the herding parameter below a critical
value, but the financial crashes
yields an increase in which the probability of large returns exists
for the herding parameter larger than the critical value.
Futhermore, it is well known that the distribution of normalized returns has
almost the form of the fat-tailed distribution $[14]$ and
a crossover toward the Gaussian distribution in financial markets.

The theoretical and numerical analyses
for the volume of bond futures transacted at Korean futures exchange market
have presented in the previous work $[15]$.
This treated mainly with the number of transactions for two
different delivery dates and found
the decay functions for survival probabilities $[16]$ in the Korean bond futures.
We also argued the tick dynamical behavior of the Korean bond futures price using
the range over standard deviation or the R/S analysis in the futures exchange market $[17]$.
In a recent paper $[18]$, Skjeltorp has shown that there exists the
persistence caused by long-memory in the time series on Norwegian and US stock markets.
The numerical analyses based on multifractal Hurst exponent and the price-price correlation
function have used for the long-run memory effects. It is
found that the form of the probability distribution of the normalized return
leads to the Lorentz distribution rather than the Gaussian distribution $[17]$.
In this paper, we investigate the dynamical herding behavior 
for the yen-dollar exchange rate in Japanese financial market.
Our obtained result will compared with other numerical calculation.

First, we introduce the yen-dollar exchange rate for two delivery dates of tick data: 
One analyzes minutely tick data for the period 1st March - 8th March 2002, 
while the other analyzes daily tick data for the period 4th January 1971 - 30th June 2003. 
We show minutely time series of the yen-dollar exchange rate $p(t)$ in Fig.$1$, and 
the price return $ R (t )$ is defined as
\begin{equation}
R (t ) =  \ln \frac{p(t + \tau )} { p(t )} ,
\label{eq:a1}
\end{equation}
where $\tau$ denotes the time interval, i.e., our average time between ticks,
and Fig.$2$ plots the price return at $\tau=$ one minute.

To describe the averaged distribution of cluster, let us suppose three return states composed by $N$
agents, i.e., the continuous tick data of the yen-dollar exchange rate. 
We assume that from the states of agent $n$ composed of the three
states $\phi_n = \lbrace -1, 0, 1 \rbrace $ the state of clusters is given by 
\begin{equation}
s(t ) =   \sum_{n=1}^{N} \phi_n ,
\label{eq:b1}
\end{equation}
where $\phi_n =0 $ is the waiting state that occurs no transactions or gets no return
and $\phi_n =1$ ($\phi_n =-1$) is the selling (buying) states,
i.e., the active states of transaction.
Assuming that it belongs to the same cluster between a group of agents sharing the
same information and making a common decision,
the active states of transaction can be represented by vertices
in a network having links of time series.

Since the distribution of returns is directly related to the distribution of
cluster, we can introduce the averaged distribution of cluster $P(s)$ 
that scales a power law 
\begin{equation}
P(s) \simeq s^{-\alpha}
\label{eq:bb12}
\end{equation}
with the scaling exponent $\alpha$. Here $P(s)$ means the probability of the cluster $s$ for selling and buying herds.
Accordingly, Fig.$3$ presents the log-log plot of the averaged
distribution of cluster against the size $s$ of the minutely
transacted states, and the values of the scaling exponent $\alpha$ are summarized in Table $1$.

The charateristic feature of herds can be well described by the probability distribution
of returns $P(R)$.
To find the distribution of the price return $R$ for different herding probabilities,
the herding parameter of the network of agents can be estimated from $P(\phi_n =+1, -1)=$$a=a_{+} + a_{-} $,
where $P(\phi_n =+1)=$$a_{+}$ and $P(\phi_n =-1)=$$ a_{-}$ are, respectively, 
the probability of the selling and buying herd.
Here we can perform the simulation of $P(R)$ by using the the herding parameter $ H \equiv (1-a)/a$, i.e.,
the ratio of no herds to a herd.  
The herding parameter is incorporated into 
the price return, whose elements are usually the random numbers proportional to the 
real data in Fig.$2$.
As it analyzes minutely and daily trading data on yen-dollar exchange rate,
the probability distribution of returns $P(R)$ for three herding parameters
satisfies the power law
\begin{equation}
 P(R) \simeq R^{-\beta}
\label{eq:c1}
\end{equation}
with the scaling exponents $ \beta=3.11$(the time interval $\tau=$ one minute) and $3.36$($\tau=$ one day) in Figs.$4$ and $5$.
The exponents for the won-dollar exchange rate and the KOSPI are, respectively, $\beta=2.2$ and $2.4$($\tau=$ one day),
as shown in Figs.$4$(a) and (b) of Ref. $[19]$.
%

Lastly, let us calculate the distribution of normalized returns.
Then, the normalized return $R(t)$ can be represented in terms of
\begin{equation}
  R(t) =(R-<R>)/\sigma, 
\label{eq:d1}
\end{equation}
where the statistical quantity $<R>$ is the value of returns
averaged over the time series of $R$
and the volatility $\sigma$ is defined as $\sigma= (<R^2 > - <R>^2 )^{1/2}$.
As the time step takes the larger value, the probability distribution as a function of 
$R(t)$ is expected to approach to the Gaussian form.
%
%
In Fig.$6$, we show
the semi-log plot of the probability distribution of the normalized returns
for the herding parameter $a=0.3$($H=2.99$),
where the time steps are taken as $\Delta t=1$ minute and $1$ day.
Here the form of the fat-tailed distribution of price returns
appears for the distribution using $\Delta t=1$ minute,
and the probability distribution of normalized returns really
reduces to a Gaussian form at $\Delta t=1$ day. 

In summary, we have investigated the dynamical herding behavior 
for the yen-dollar exchange rate in Japanese financial market.
Specially, the distribution of the price return scales as a power law
$R^{-\beta}$ with the exponents $ \beta=3.11$($\tau=$ one minute) and $3.36$($\tau=$ one day),
while the scaling exponent shows $ \beta=2.2$ for the won-dollar exchange rate$[19]$.
However, our distributions of the price return are
not in good agreement with the other result$[5]$.
It is in practice found that our scaling exponents $ \beta $ are somewhat larger than the
numerical $1.5$.
It would be noted that the probability existing financial crashes
is high, because the active herding behavior
occurs with the increasing probability as the herding parameter
becomes larger value in real financial Markets.

We can find the critical herding parameter $H^{*} =2.33$$(a=0.3)$ from our minutely tick data,
similar to the case for the won-dollar exchange rate and the KOSPI.
The critical herding parameter $H^{*}$ denotes the crossover value for financial crashes of selling and buying states.
It is obtained that the financial crashes occur at $a<0.3$($H>2.33$).
In Japanese exchange market, it occurs no crash for daily tick data from Fig.$5$, while
the crash regime appears to increase in the probability of high returns values 
for minutely tick data. 
It is also found that the distribution of normalized returns 
reduces to a Gaussian form, 
and there arises a crossover toward a Gaussian probability function
for the distribution of normalized returns.
Our analyses plan to investigate in detail the herd behavior
for other foreign exchange rates and the auction items $[20]$, and
we also hope that the dynamical herd behaviors apply to many tick data
in financial markets.\\

\vskip 5mm
\noindent
{\bf ACKNOWLEDGMENT}
\vskip 2mm
\hfill\\
This work was supported by grant No.R01-2000-000-00061-0 from the Basic Research 
Program of the Korea Science and Engineering Foundation.\\

%
%
\newpage
\vskip 10mm
\begin{center}
{\bf FIGURE  CAPTIONS}
\end{center}

\vspace {5mm}

\noindent
Fig. $1$.  Time series of the yen-dollar exchange rate $p(t)$.
\vspace {10mm}

\noindent
Fig. $2$.  Plot of the price return $R_{\tau=1} (t ) = $ ln$[p(t +1 )/p(t )]$ for
minutely tick data.
\vspace {10mm}

\noindent
Fig. $3$.  Plot of the averaged probability distribution of cluster sizes $s$
for the herding probability
$a=0.3$ ($H=2.33$), where the averaged probability distributions for the yen-dollar exchange rate
scales as a power law $s^{-\alpha}$ with the exponent $ \alpha=4.64$(the dot line).
\vspace {10mm}

\noindent
Fig. $4$.  Log-log plot of the probability distribution of returns for three types of herding probabilities
$a=0.1$, $0.3$, $0.5$ ($H=9$, $2.33$, $1$), where the dot line scales as a power law
$R^{-\beta}$ with the exponent $3.11$($\tau=$ one minute).

\vspace {10mm}

\noindent
Fig. $5$.  Log-log plot of the probability distribution of returns for three types of herding probabilities
$a=0.1$, $0.3$, $0.5$, where the dot line scales as a power law
$R^{-\beta}$ with the exponent $3.36$($\tau=$ one day).

\vspace {10mm}

\noindent
Fig. $6$.  Semi-log plot of the probability distribution of the normalized returns
for herding probability $a=0.3$($H=2.33$),
where the dot line is the form of Gaussian function with $\sigma=7.1$ for
the yen-dollar exchange rate. 

\vskip 10mm

\begin{center}
{\bf TABLE  CAPTIONS}
\end{center}
\vspace {5mm}
\noindent
Table $1$.  Summary of values of the scaling exponent $\alpha$
for the yen-dollar exchange rate.
\vspace {3mm}\\
\begin{tabular}{lcr}\hline\hline
    $\tau$                  &  $a$                      &  $\alpha$  \\  \hline                           
                            &  $0.1$                    &  5.44         \\
   one minute               &  $0.3$                    &  4.64         \\ 
                            &  $0.5$                    &  4.39         \\  \hline
                            &  $0.1$                    &  5.49         \\
  one day                   &  $0.3$                    &  4.97         \\
                            &  $0.5$                    &  4.02         \\  \hline \hline
\end{tabular}

\end{document}